\documentclass[]{spie}  

\usepackage{amsmath,amsfonts,amssymb}
\usepackage{graphicx}
\usepackage[colorlinks=true, allcolors=blue]{hyperref}
\usepackage{subcaption}
\title{Integration of Data Reduction and Near Real-Time Archiving into the Keck Observing Model}

\author[a]{Max N. Brodheim}
\author[a]{John O'Meara}
\author[a]{Jeffrey A. Mader}
\author[b]{G. Bruce Berriman}
\author[a]{Matthew Brown}
\author[a]{Lucas Furhman}
\author[a]{Tyler Tucker}
\author[b]{Christopher R. Gelino}
\author[b]{Meca S. Lynn}
\author[b]{Melanie A. Swain}
\affil[a]{W. M. Keck Observatory, Hawai'i, United States}
\affil[b]{Caltech/IPAC-NExScI, Pasadena, United States}

\begin{document} 
\maketitle

\begin{abstract}

The W. M. Keck Observatory is welcoming a new era where data reduction and archiving are tightly integrated into our observing model, under the auspices of the Observatory's Data Services Initiative (DSI) project. While previously the Keck Observatory Archive (KOA) archived minimally processed, raw science data the day after observing, Keck is transitioning to a model in which it archives both raw frames and reduced data in near real-time. These data will be made available to observers and collaborators immediately upon ingestion through a dedicated new interface that will support collaboration and sharing among teams, as well as stream data directly to personal computers without access to Keck's internal networks. Both the raw and science-ready data products will be made publicly available upon the expiration of data protections.

The Keck Cosmic Web Imager (KCWI) instrument is the first whose data are managed this way. It showcases how KOA integrates into an observing night, provides the data needed to make real-time adjustments to observing, and delivers products that allow for faster publication by both our observers and archival researchers. This effort has involved the delivery of new, compact, Python-based data preparation and ingestion software.

We also discuss the new and updated Data Reduction Pipelines (DRPs) required to generate science-ready data, how their development and deployment enables the delivery of these products, and how Keck's commitment to maintaining DRPs in-house will result in more robust datasets for all our observers and KOA users.

\end{abstract}


\section{INTRODUCTION}
\label{sec:intro}  

The W. M. Keck Observatory's (WMKO) observing model has remained generally consistent since first light in 1990: observers are assigned time, conduct their observations, and receive their raw data at some point after observing concludes. Although WMKO has streamlined each of these steps with steady improvements as new technologies become available, the core model has never been updated. For example, although current users can retrieve their observations via an internet connection as opposed to writing physical media at the end of observing (as was the case for much of WMKO's history), they still typically retrieve their data only after the night has ended. While this  model has provided top-tier data for multiple decades, there are several areas where WMKO can and should evolve, in order to promote observatory efficiency and improve the observer experience, while at the same time decreasing the amount of time between observation and publication. This document focuses on changes being made to the data reduction and retrieval experience as part of WMKO's Data Services Initiative (DSI), a NASA funded project aiming to improve the accessibility and usability of data collected at Keck.\footnote{For more information on other changes to the operational model at WMKO and DSI, see \textit{The W.M. Keck Observatory Data Services Initiative: Creating data that is useful, usable, and quick} (J. O'Meara, SPIE 12186) and \textit{Modernizing Observation Planning For Accessible, Science-Ready Data} (M. Brown, SPIE 12189)} The work described was conducted as a joint effort between WMKO, IPAC/NExScI (the developers and hosts of the Keck Observatory Archive), and other software partners.

\section{Goals}
\label{sec:goals}

One source of observing inefficiency stems from the difficulty observers can face while trying to determine their data's scientific viability during their observing run. For example, while some observers' needs are satisfied by simple alignment checks or visual confirmations that their target is in frame, others might need to know the signal-to-noise ratio of a specific stellar emission line in order to judge whether their current observing plan will meet their scientific requirements. In the worst cases, this can vastly increase time to publication, since observers may require unexpected follow-up nights to collect new data, an expensive and time consuming process.

Assuming that good data are collected, the process of retrieving those data from WMKO is not optimized. Prior to the work described in this document, WMKO observers were offered two options: use the Keck Observatory Archive (KOA) web interface, which displayed data only once they were archived hours later, or transfer their data directly from Keck's servers to their own machine using command line tools like rsync. Both of these options did not provide optimal functionality to users: using KOA required waiting until the morning, which eliminates all chances of middle-of-the-night reductions, and rsyncing files exposes Keck's internal data servers to all users, with little room for restrictions. Once observers do collect their data, it then often takes a long time to generate the reduced data products that are required for research and eventual publication. These inefficiencies arise from several root causes. 

First and foremost, WMKO had not offered formal support for data reduction pipelines (DRPs) until the initiation of the DSI project, with limited exceptions. While all of our instruments have at least one DRP associated with them, they vary vastly in capability, reliability, and usability. For example, some require knowledge of and access to the proprietary programming language IDL (e.g. the OSIRIS DRP), and some support only specific science cases and instrument configurations (e.g. the DEIMOS DEEP2 DRP). 

To add to the confusion, some instruments have several community-made DRPs that all do subtly different things, which requires users to navigate several options on their path towards reducing their data. As spectrograph instruments and their corresponding data products get more complex, the lack of supported DRPs will only hinder WMKO users more. This lack of DRP support also directly impacts observing: there is currently no reliable means to determine whether spectral data collected during the night are scientifically viable without users running DRPs on their own.

Secondly, the complexity of Keck's instrument suite generally requires researchers to have a deep understanding of the instruments they use in order to extract the equipment's full potential. While this is not a fundamental issue to the operational model at Keck, it does disadvantage anyone hoping to use data from KOA to conduct archive-based research.

Finally, there is no easy way to share Keck data effectively between observing team members. At the moment, the only way for observers to share data with their team is to use external file sharing tools such as Dropbox, transfer data manually from the command line, or exchange physical media.

In light of these areas for improvement, DSI has set several goals and requirements for WMKO. These goals broadly outline a path towards providing observers their data, both raw and reduced, quickly and easily. Additionally they emphasize the importance of providing well documented data reduction tools to all users. The full list of goals and requirements are presented in Appendix \ref{ap:goals}.

\section{Approaches to Automated Data Reduction and Dissemination at Classically Scheduled Observatories}
\label{sec:approach}

WMKO is a classically scheduled observatory, which means that astronomers conduct their observations themselves on an assigned night. Classical scheduling exists as the counterpart to queue-based observing, in which an observatory staff member carries out observations on behalf of astronomers whenever specified observing constraints are met. The processes and infrastructure required to reduce and share data collected at an observatory typically differ substantially between queue and classical observatories. This section overviews the various approaches that WMKO and KOA considered towards data reduction and archiving, as well as our choices and justifications.

\subsection{DRP Development and Support}

In an ideal world, WMKO would develop all DRPs in-house, with guidance and development support from instrument design teams. This would allow WMKO to ensure all DRPs meet our requirements for usability and quality, while also ensuring long-term support for all our instruments. This also would allow WMKO to ensure that each DRP meets the requirements outlined in Section \ref{sec:goals}. However, this approach requires both substantial resources and expertise that are not currently within the scope or capability of the observatory.

An alternative is to continue on the path that WMKO previously followed, leaving DRP development and maintenance entirely to the observing community. This relies on our observers to develop the pipelines they need, leveraging their expertise in the process. However, this approach only guarantees results for only those individual developers and their teams, who are incentivized to develop features they need for their research, as opposed to common community-use functionality. Furthermore, it does not ensure good documentation, support, or an avenue for open-source contribution.

Finally, WMKO can hybridize the approaches above, such that WMKO staff participate in the administration of DRP development, while drawing upon instrument teams and the open-source community for contribution. This takes advantage of the experience present in our observing community, while allowing WMKO to ensure that all DRPs are of sufficient quality to meet the needs of both observers and WMKO. 

WMKO has elected to adopt the final approach above for a number of reasons. Given the number of instruments at the observatory, implementing custom DRPs for each is not realistic in a short time frame. This hybrid approach takes the form of two initiatives: one aimed at existing instrumentation, and one at future instruments. 

WMKO has partnered with the PypeIt project to provide DRP support for all spectrograph instruments delivered prior to KCWI (due to their relative simplicity, imaging DRPs are typically more accessible and general purpose, and therefore less critical for WMKO to develop). PypeIt is a Python package that provides semi-automated reductions of spectral astronomical data. \cite{pypeit:joss_pub} While it supports spectrographs from several observatories (and can be easily upgraded to support more) WMKO is providing support for the project in return for a focused development of PypeIt's capabilities with the WMKO instrument suite. PypeIt currently offers full support for DEIMOS and MOSFIRE data in all observing modes, and is currently implementing support for reducing LRIS data. In order to ensure consistency in DRPs for future instrumentation, WMKO has also mandated that all new DRPs implement the Keck DRP Framework. This is a Python package, designed and maintained by WMKO, that provides a modular task based queuing framework for DRP creation and implementation.\footnote{For more information, see Section \ref{kcwi:drp} and  \url{https://keckdrpframework.readthedocs.io/en/latest/index.html}}\cite{rizzi_kwok_brown}

\subsection{Data Dissemination}

WMKO currently has two primary methods for sharing data with observers: offering low-level access directly to the disks that store the raw data at the observatory, or downloading data the morning after a night from KOA. The former method is easy for observers, but requires WMKO to expose its internal network to outside users, and limits its ability to restrict access to other observer's proprietary data. The latter method does not allow users to access their data in near real-time, which renders the generation of quick-look data products impractical. In order to continue to provide users with quick and easy access to their data, while also ensuring proprietary protections for other users and easy searchability, WMKO and KOA have endeavoured to update the archive to support near real-time ingestion of data.

\section{Progress So Far}
\label{sec:progress}

\subsection{Data Reduction Pipelines}

Through DSI, WMKO and KOA have made major steps towards providing both DRPs and reduced data products to our users. As a requirement, all DRPs provided by WMKO (both PypeIt led and otherwise) will be delivered with a quick-look mode. This mode typically takes the form of lighter restriction on fitting routines, skipping certain calibration steps, and specifying specific spectral orders or mask-slits for examination as opposed to entire frames. These quick-look data are only as useful as they are available to users, so WMKO will automatically run these pipelines on-site, providing their outputs to users as quickly as possible. WMKO currently carries out quick-look reductions for KCWI, and is currently developing an interface to provide those products to all instruments supported by PypeIt.

WMKO has also begun embedding a DSI team member in all future instrument DRP development teams. This allows WMKO to steer development towards providing the functionality required by Section \ref{sec:goals}, while also organically growing WMKO expertise with the DRPs that it will be supporting in the future. Currently, that embedded support is present for three incoming instruments/upgrades: the Keck Cosmic Reionization Mapper (an upgrade to KCWI), the Keck Planet Finder (an ultra-stable high resolution spectrograph), and the Keck Planetary Imager and Characterizer (an exoplanet imaging system).

\subsection{Real Time Ingestion}
\label{sec:rti}

DSI has developed Real-Time Ingestion (RTI) as the primary way to improve user experience with regards to receiving their data. RTI is comprised of several connecting pieces: software that monitors WMKO's data directories for new files and adds headers that are required for archiving, an upgraded file transfer system to stream data on a file-by-file basis, and a newly upgraded interface to access those files. Over the last three years, several improvements have been made to WMKO's data processing infrastructure to support the transition of RTI. 

	\begin{figure} [ht]
	\begin{center}
		\begin{tabular}{c} 
			\includegraphics[height=7cm]{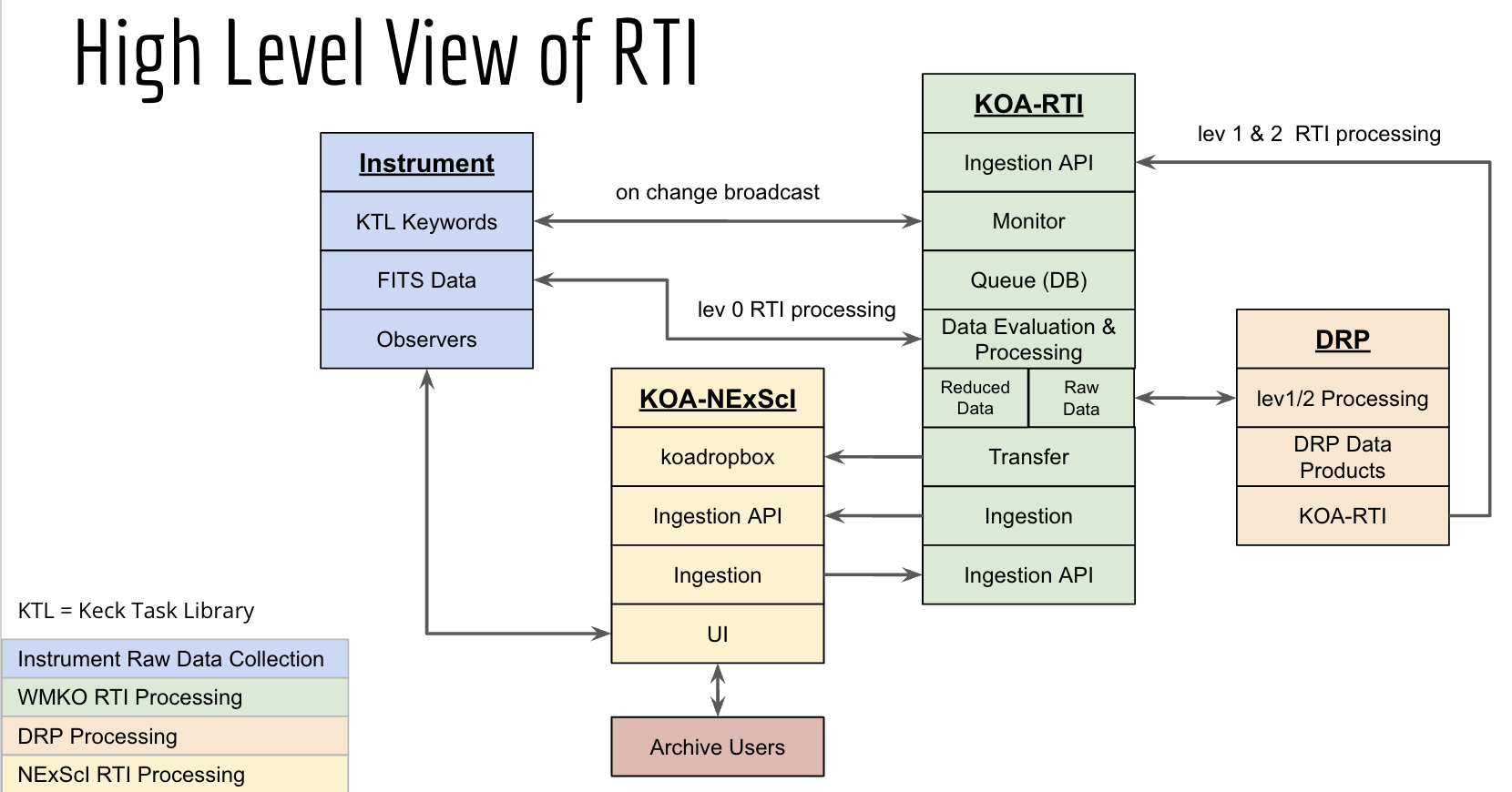}
		\end{tabular}
	\end{center}
	\caption[example]{Flow chart describing the various components of RTI.} 
	{ \label{fig:flowchart} }
\end{figure} 

Prior to RTI, data ingestion at WMKO was conducted by a suite of custom procedures written in IDL, running at scheduled intervals. This software was ported to Python 3 and re-written to utilize object-oriented programming techniques. The new software is called to execute by an improved monitoring system based directly off of KTL\footnote{Keck Task Library, or KTL, is the keyword-based control system used at Keck to execute virtually all observatory functions}, as opposed to scheduling scripts or file system related triggers. This KTL monitor uses callbacks built into the KTL system to begin data processing and the archiving of FITS files on an individual basis. This lower-level triggering makes data ingestion both faster and more reliable, easily handling calibration and engineering frames collected at unusual times of day. Data ingestion at KOA is carried out by a recently re-implemented, compact software package written in C, which uses highly-modular code to provide support for all current and future instruments using the current KOA archive configurations.

In addition to the improvements to data processing and ingestion at internally at Keck, RTI has implemented a robust HTTP-based system to facilitate communication (both file transfers and ingestion triggers) between WMKO in Hawai'i and IPAC/NExSCI in Pasadena. This new system allows WMKO to initiate the transfer of individual files, instead of informing KOA via email of a batch of files ready to be transferred.

To tie all of this back-end work together for the user, the Real-Time User Interface (RTUI) will be deployed. RTUI is a web-based system that allows users to monitor, examine, and download both raw and reduced data throughout the night. The UI uses WebSockets to communicate with user clients, allowing multiple authorized users to view data at the same time. RTUI will also allows users to download files on an individual one-off basis, as well as opening a stream to download all files locally as soon as they become available.

\subsection{KCWI as a Case Study}

WMKO has, as described above, made progress towards accomplishing its data reduction and dissemination goals. While substantial steps forward have been made for many of our instruments, so far the only fully upgraded instrument system at the observatory is the Keck Cosmic Web Imager (KCWI), an IFU spectrograph. This upgrade included several upgrades to the KCWI DRP, along with deploying and integrating RTI into observatory operations.

\subsubsection{Upgrading the KCWI DRP}
\label{kcwi:drp}

\begin{figure}
    \centering
    \includegraphics[width=10cm]{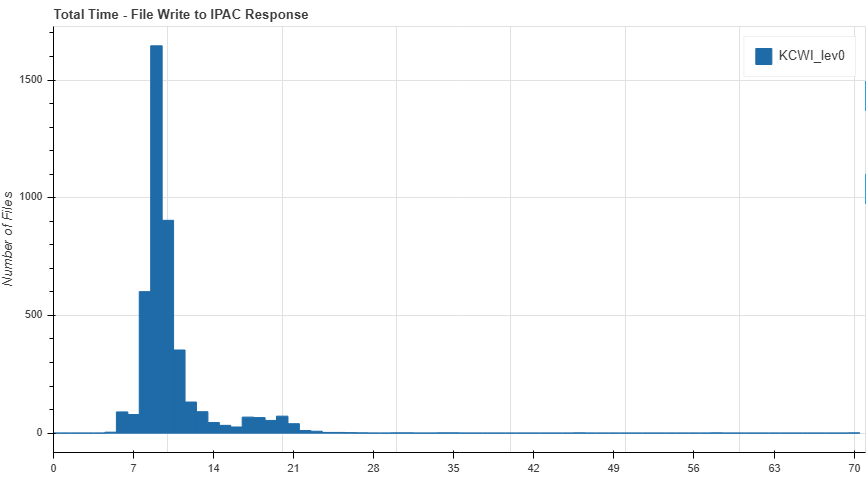}
    \includegraphics[width=10cm]{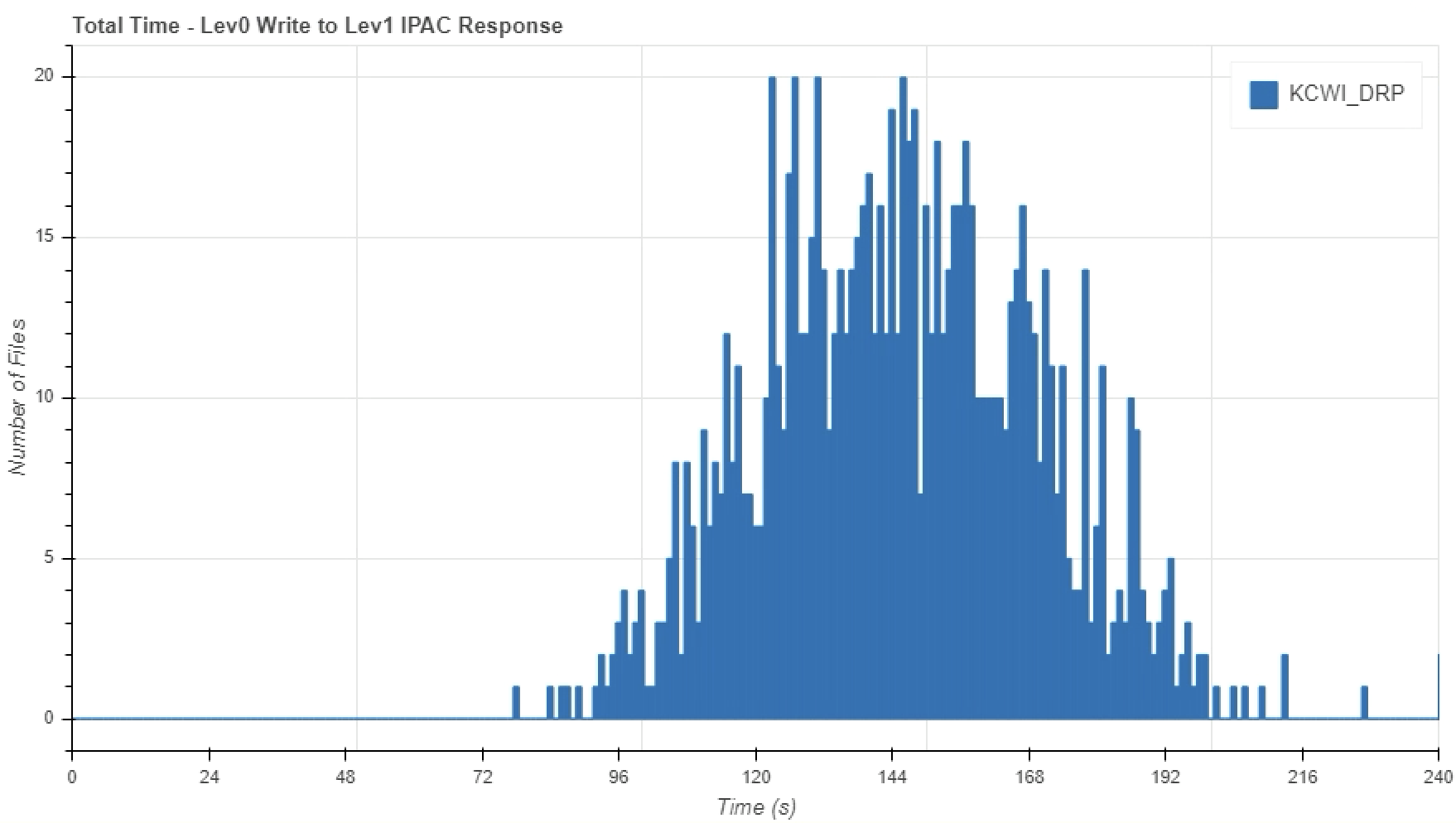}
    \caption{Histograms showing the time between file write-to-disk and archiving at KOA, for both raw (top) and quick-look (bottom) data. KCWI offers users two detector readout modes that yield different file sizes, which is reflected in the two peaks present in the level 0 chart. Since the vast majority of KCWI users utilize the smaller file size, the first peak is larger. Both the raw and quick-look data ingestion times fall well within their respective requirements.}
    \label{fig:ingest_times}
\end{figure}

While KCWI was delivered to WMKO with a functioning DRP, several changes were needed to bring the pipeline up to the standards required for the project described in this paper.\cite{Morrissey_2018} These changes included porting the pipeline from IDL to Python using the Keck Data Reduction Pipeline Framework, creating a quick-look mode, and developing the infrastructure needed to run the pipeline automatically, along with implementing general bug fixes and new capabilities.

Much of the above work was accomplished through the use of Keck's DRP Framework. This package, now required to be used by all new WMKO instrument DRPs, provides a framework that helps developers design and implement automated pipelines.\cite{rizzi_kwok_brown} It provides utilities for running a pipeline in a watchdog-like mode, generating different pipeline versions to handle varying user needs, along with other general pipeline functionality. Usage of this framework implemented by default all of the infrastructure upgrades required, including directory monitoring and automated ingestion and reduction of data into the pipeline.

The quick-look mode of the pipeline proved to be straightforward to implement: once all calibrations have been reduced (excepting stellar standard frames), the pipeline automatically generates master calibrations files, and uses them to fully reduce each science file as they are collected. Usage of multiprocessing and a server that is not resource-constrained means that a full scientific reduction can be obtained within the time limits set out in Section \ref{sec:goals}. The only difference between the quick-look and full science modes of the pipeline are whether frames are flux calibrated, since observers often collect the standard frames needed to compute the sensitivity function after they collect their science frames.

The new Python KCWI DRP was released in June 2021, and receives regular updates from a combination of the open-source community, WMKO developers, and KCWI development team members.\cite{KCWI_DRP_IPAC} As seen in Figure \ref{fig:ingest_times}, all quick-look products are being generated and archived within the time frame specified in Appendix \ref{ap:goals}. Science-ready data products are created and transferred to KOA on a similar time-scale, although those reductions are not triggered to begin until the following morning. These reductions also occur well before the time limits required.

\subsubsection{Upgrading RTI for KCWI}

\begin{figure} [ht]
   \begin{center}
   \begin{tabular}{c} 
   \includegraphics[width=15cm]{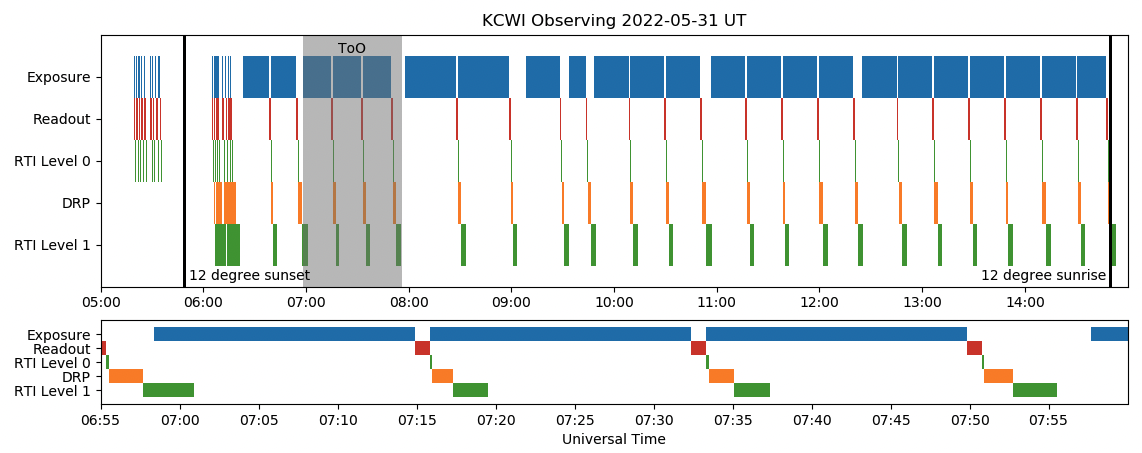}
	\end{tabular}
	\end{center}
   \caption[example]{Time required by various KCWI RTI functions from a typical night of observing, displaying a whole night (top) and a magnified view of the highlighted section (bottom). Level 0 refers to raw data archiving time, DRP to the amount of time the pipeline took to reduce each frame, and Level 1 to quick-look data archiving time. Exposure and readout still dictate the cadence of observing, with RTI processes happening simultaneously.} 
   { \label{fig:time_series} }
\end{figure}

Although most WMKO instruments now use RTI to archive data, KCWI's full DRP integration allows a detailed analysis of RTI capabilities. As seen in Figure \ref{fig:ingest_times}, KCWI raw data are typically archived within 8 seconds of files being written to disk, with some variation arising from different file sizes and networking speeds between Hawai'i and California. As soon as each file is processed by RTI, it is handed off to the KCWI DRP and reduced, creating quick-look data products, as seen in Figure \ref{fig:time_series}. At the conclusion of a night of observations, the KCWI DRP is triggered again, this time reducing all files, including any calibrations that were collected after sunrise. These reduced data are archived immediately upon the termination of the DRP task.

\subsection{Results and Future Steps}

\begin{figure}
    \centering
    \includegraphics[width=10cm]{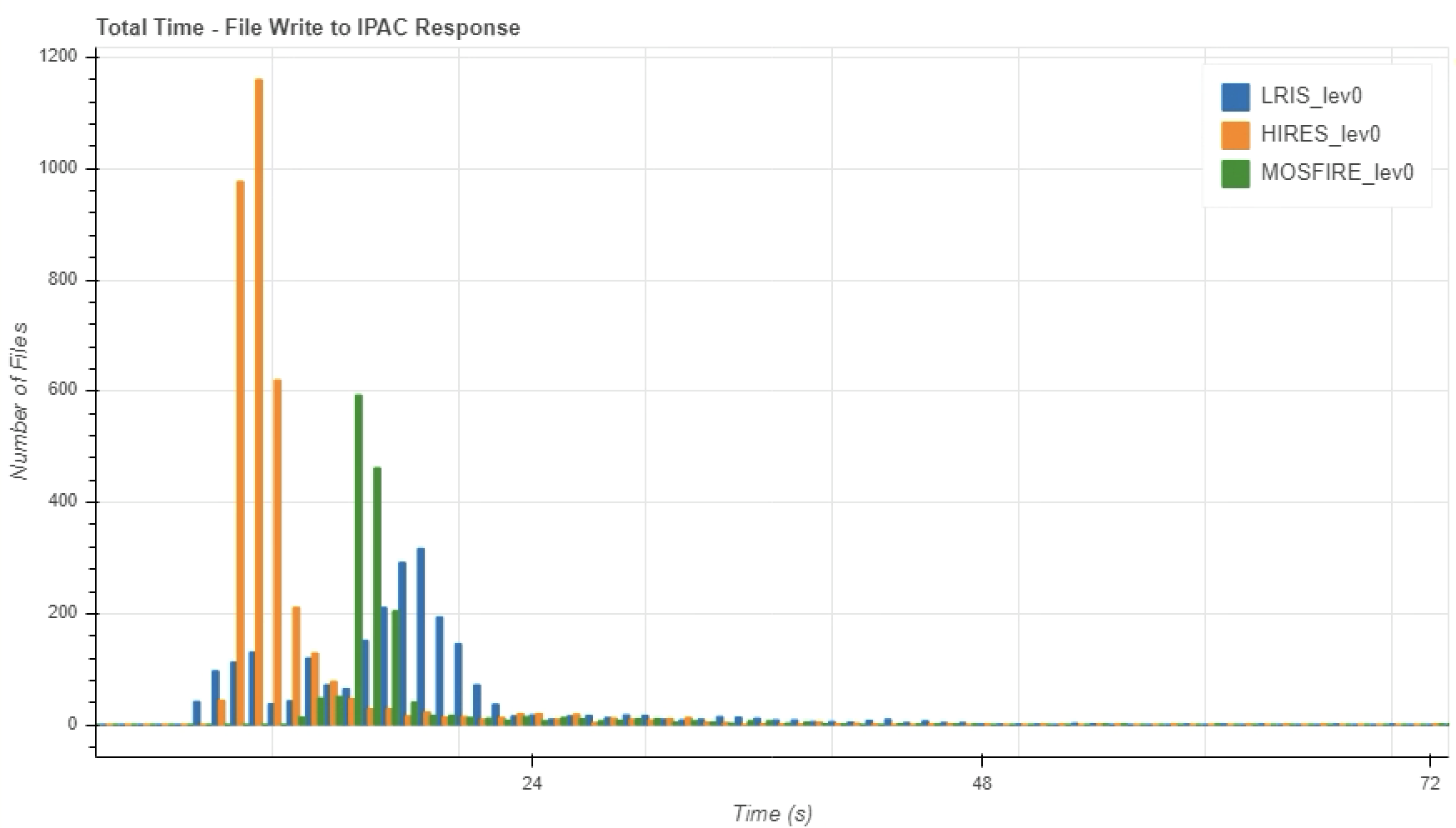}
    \hfill
    \includegraphics[width=10cm]{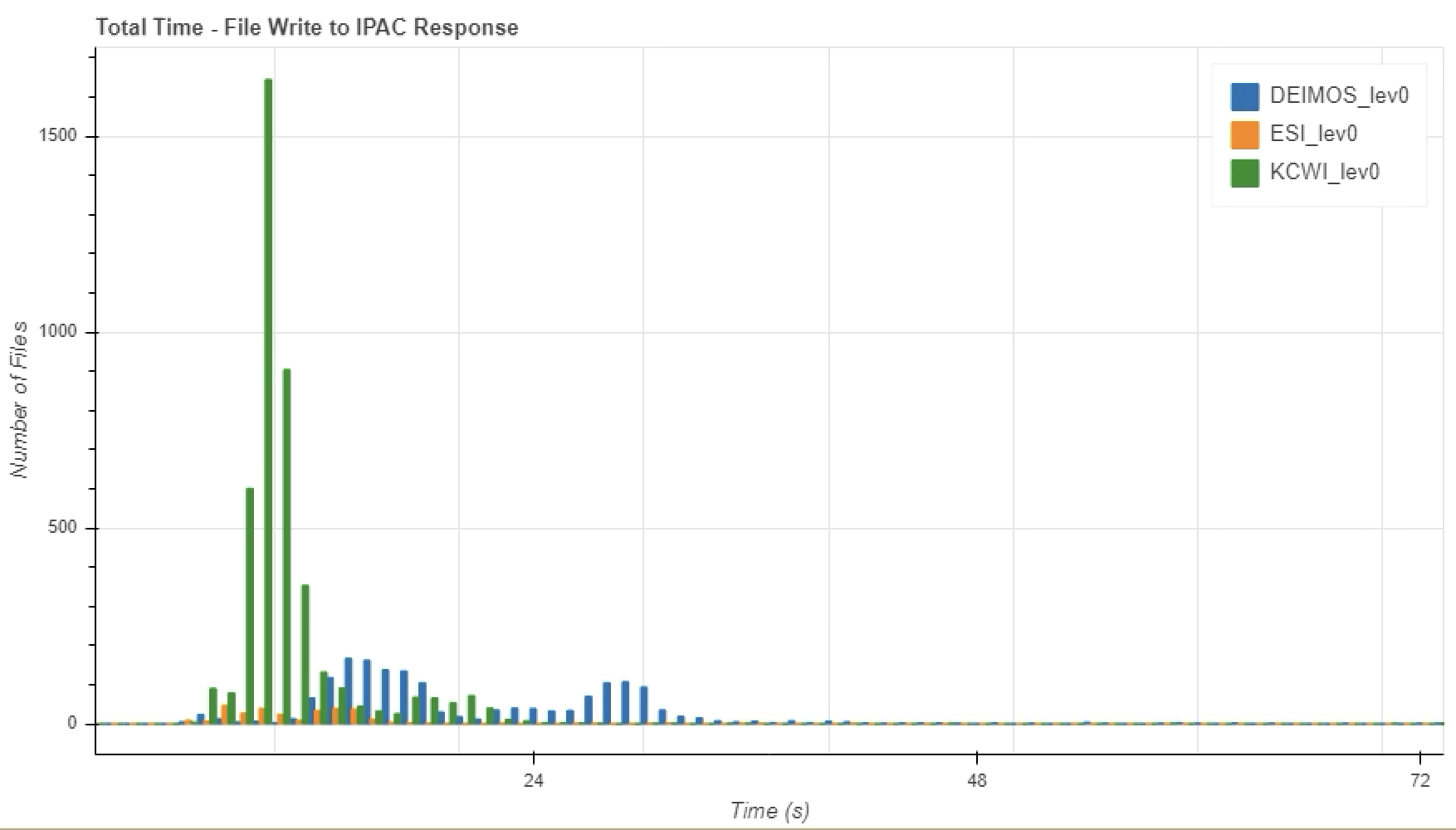}
    \caption{Raw data ingestion times for Keck I (top) and Keck II (bottom), for all RTI enabled instruments. Ingestion times typically rest under one minute, while the requirements specify a maximum of five.}
    \label{fig:all_ingest}
\end{figure}

While it is difficult to generate quantitative findings about the success of the DRP work described above, there are some general conclusions to be drawn, along with some benchmarks to look out for in anticipation of future data. RTI has easily surpassed all timing requirements laid out in Appendix \ref{ap:goals} for all instruments it has been deployed for, as seen in Figure \ref{fig:all_ingest}.

The upgrade of the KCWI DRP has been largely successful. Although re-writing the code in Python introduced some small errors, overall the DRP runs faster, is better documented, and more feature-rich than it was before. The use of the DRP Framework enabled rapid deployment of the automated version of the pipeline, and critically lays the foundation for the deployment of all future DRP Framework-based pipelines at WMKO. It is expected that future ports of IDL pipelines to Python (such as the OSIRIS DRP) will yield similar results.

\begin{figure} [ht]
   \begin{center}
   \begin{tabular}{c} 
   \includegraphics[width=15cm]{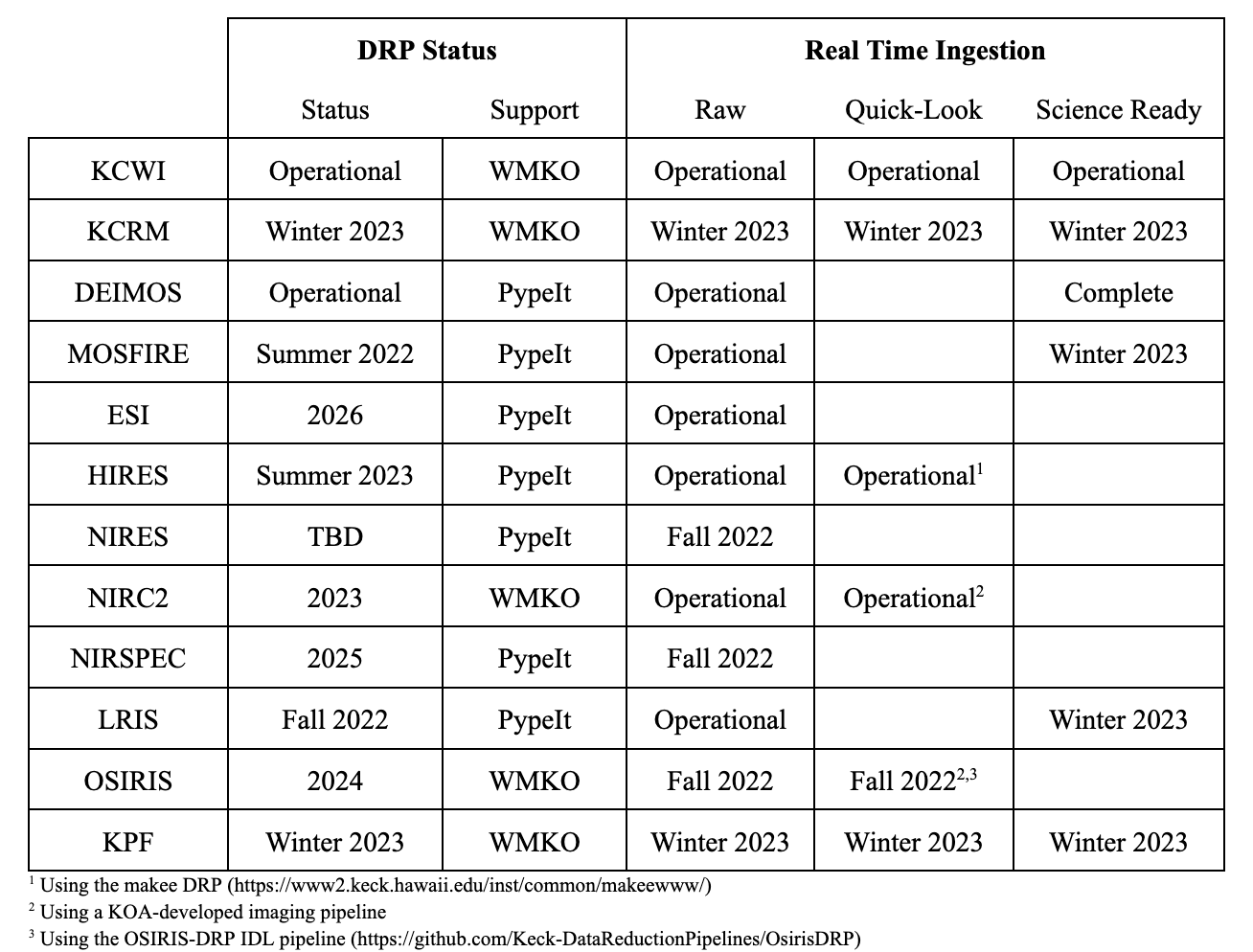}
	\end{tabular}
	\end{center}
   \caption{Table describing the status of DRP and RTI development/deployment at WMKO. Instruments without entries are planned, but not scheduled.} 
   { \label{tab:dates} }
\end{figure} 

All of the infrastructure built to transfer data from the pipeline to RTI for archiving is compatible with future DRPs, be they based on PypeIt or the DRP Framework. As soon as a DRP is completed to WMKO standards, it can be easily integrated into RTI and deployed to operations. Table \ref{tab:dates} overviews the current schedule for DRP and RTI development at WMKO.

The implementation of RTI at WMKO has been largely successful, for a number of reasons. Troubleshooting can now be carried out on a file-by-file basis, as opposed to batch-reprocessing an entire night's data. This decreases data archiving's load on WMKO's network, speeds up response time to errors, and allows for on-the-fly adjustments to RTI infrastructure when required. Prior to RTI's deployment, a warning during ingestion of one file would require the entire system to be turned off, fixed, and restarted, possibly missing files in the interim.

Next, although the swap from email-based communication to HTTP was initially intended only for data transfer between WMKO and IPAC/NExScI, having an API based ingestion system proved to be helpful when integrating new functionality into the observing experience. For example, no substantial infrastructure changes were needed to begin ingesting reduced DRP products, as the DRPs can interact directly with the HTTP API when files are ready for archiving. Another unexpected benefit has been the collection of detailed networking information that was previously not logged. Since RTI touches several different systems at WMKO, and communicates with IPAC/NExScI, it provides a unique viewpoint to diagnose network slowdowns, identify network outages, and generally log useful infrastructure data that previously would have been missed.

Additionally, port of RTI functionality from IDL to Python generally improved code clarity and exposed WMKO to standard programming practices used in industry. It is expected that further development will continue to adhere to industry standards, yielding more reliable, maintainable, and efficient products.

Along with the successes in both DRP development and deployment come some lessons learned. The use of the DRP Framework has proven to be a sticking point for some DRP teams, as new instrumentation sometimes come with novel data flows which can be challenging to adapt to a task-queue based framework. Further development time and resources need to be allocated to keep the framework up-to-date with developer needs to fully maximize the advantages that it can bring to WMKO and the scientific community.

Similarly, end-user feedback on RTI's various interfaces and use-cases should have been collected earlier and more frequently. While users were polled on their preferences, and that feedback was used to inform design decisions, increasing engagement with users throughout the development process yields both a better end product and increased user buy-in.

\section{Conclusions}
The steps taken in both data reduction and near real-time archiving demonstrate WMKO and KOA's first steps towards an improved observing model at Keck. While more work is required with regards to DRP development and RTI UIs, user feedback has been generally positive and the progress made so far indicate that WMKO's commitment to streamlining the observing experience is on track to be successful.

\appendix
\section{Goals and Requirements}
\label{ap:goals}
\begin{enumerate}
	\item Provide our users with near real-time access to their data remotely, without access to WMKO's network, and provide those users a way to share that data with others within KOA, to support both traditional and Time-Domain Astronomy (TDA) observers
	\begin{enumerate}
		\item Observers shall have access to lev0 (raw) data through the KOA web interface within 5 minutes of instrument readout
		\item Observers shall have access to lev2 (fully reduced) data products through the KOA web interface within 24 hours of instrument readout
	\end{enumerate}
	\item Provide our users quick-look data that allows observers the ability to determine scientific viability with enough time to make changes to their observing program
	\begin{enumerate}
		\item Observers shall have access to lev1 (quick-look) data products through the KOA web interface within 5 minutes of DRP completion
	\end{enumerate}
	\item Provide our users, both observers and archival researchers, access to robust, well documented, and actively supported DRPs
	\begin{enumerate}
		\item All WMKO-supported DRPs shall be fully open-source
		\item All future WMKO DRPs shall use the Keck Data Reduction Pipeline Framework
		\item WMKO shall verify and support a minimum of 4 instrument DRPs by February 2023
	\end{enumerate}
\end{enumerate}

There are also general goals that are hard to generate closely specific requirements for, especially in the near-term:

\begin{enumerate}
	\item Decrease the amount of time spent collected exposures that are not scientifically viable
	\item Decrease the amount of time between observations and their subsequent publication
	\item Increase the number of publications using archival data retrieved from KOA
\end{enumerate}

\acknowledgments
 
This research has made use of the Keck Observatory Archive (KOA), which is operated by the W. M. Keck Observatory and the NASA Exoplanet Science Institute (NExScI), under contract with the National Aeronautics and Space Administration. The data presented herein were obtained at the W. M. Keck Observatory, which is operated as a scientific partnership among the California Institute of Technology, the University of California, and the National Aeronautics and Space Administration. The Observatory was made possible by the generous financial support of the W. M. Keck Foundation.

The authors wish to recognize and acknowledge the very significant cultural role and reverence that the summit of Mauna Kea has always had within the indigenous Hawaiian community. We are most fortunate to have the opportunity to conduct observations from this mountain.

\bibliography{report} 
\bibliographystyle{spiebib} 

\end{document}